\documentclass[preprint,showkeys,aps,prb]{revtex4}
\usepackage[dvips]{graphicx}
\begin{document}

\title{Time-Symmetry Breaking in Hamiltonian Mechanics}

\author{
Wm. G. Hoover and Carol G. Hoover               \\
Ruby Valley Research Institute                  \\
Highway Contract 60, Box 601                    \\
Ruby Valley, Nevada 89833                       \\
}

\date{\today}


\vspace{0.1cm}

\begin{abstract}

Hamiltonian trajectories are strictly {\it time-reversible}.  Any time series of Hamiltonian
coordinates $\{ \ q \ \}$ satisfying Hamilton's motion equations will likewise satisfy
them when played ``backwards'', with the corresponding momenta changing signs :
$\{ \ +p \ \} \longrightarrow \{ \ -p \ \}$ .  Here we adopt Levesque and Verlet's
precisely {\it bit-reversible} motion algorithm to ensure that the trajectory reversibility is
{\it exact}, with the forward and backward sets of coordinates identical.  Nevertheless, the
associated instantaneous {\it Lyapunov instability}, or ``sensitive dependence on initial
conditions'' of ``chaotic'' (or ``Lyapunov unstable'') bit-reversible coordinate trajectories
can still exhibit an exponentially growing time-symmetry-breaking {\it irreversibility}
$\simeq e^{\lambda(t)}$.  Surprisingly, the positive and negative exponents, as well as the
forward and backward Lyapunov spectra ,
$\{ \ \lambda^{\rm forward}(t) \ \}$ and $ \{ \ \lambda^{\rm backward}(t) \ \}$, are usually
{\it not} closely related, and so give four differing topological measures of ``local'' chaos.  We have
demonstrated this symmetry breaking for fluid shockwaves, for free expansions, and for chaotic
molecular collisions. Here we illustrate and discuss this time-symmetry breaking for three
statistical-mechanical systems, [1] a minimal (but still chaotic) one-body ``cell model'' with a
four-dimensional phase space; [2] relatively small colliding crystallites, for which the whole Lyapunov
spectrum is accessible; [3] a near-continuum inelastic collision of two larger
400-particle balls. In the last two of these pedagogical problems the two colliding bodies coalesce.
The particles most prone to Lyapunov instability are dramatically different in the two time directions.
Thus this Lyapunov-based symmetry breaking furnishes an interesting Arrow of Time.

\end{abstract}

\maketitle

\section{Introduction}

The goal we pursue here is improved microscopic understanding of the thermodynamic irreversibility
described by the Second Law of Thermodynamics\cite{b1}.  Unlike the microscopic mechanics which
underlies it, the Second Law is strictly irreversible, and applies to macroscopic thermodynamic
descriptions of macroscopic processes in which fluctuations are ignored.  In Clausius' formulation
the Law states that the entropy of an isolated system cannot decrease.  The size of the ``isolated
system'' can be anywhere in the range from atomistic to astrophysical so long as the entropy concept
makes sense for it.  There is no reason to imagine that quantum effects or
relativistic effects or gravitational effects are crucial to the Law.  Accordingly, we limit ourselves to
{\it classical} nonrelativistic atomistic models, with short-ranged attractive and repulsive forces,
obeying Hamilton's (or, equivalently, Newton's) time-reversible equations of motion. In
particular we emphasize a many-body process for which the apparent irreversibility is especially
clearcut.  In this example two similar crystalline bodies undergo an {\it inelastic collision} in which
their kinetic energy is converted to heat.  The colliding bodies start out with minimum internal energy and
with classical entropy minus infinity.  The bodies collide and form a single oscillating liquid
drop.  Then these oscillations equilibrate.  Ultimately the equilibrated drop's internal energy is given by the
initial kinetic energy of the two colliding bodies in the frame of the full system's center of mass.

Gibbs' statistical mechanics provides the conceptual basis for thermodynamics, through Liouville's
Theorem and Hamiltonian mechanics\cite{b2}.  In that mechanics, access to all those coordinate-momentum
phase-space $\{ \ q,p \ \}$ states consistent with the initial conditions is typically provided by chaos.
Chaos is the sensitive, exponentially-growing time dependence of any small perturbation, either forward
in time, $\propto e^{\lambda t_f}$ , or backward in time,  $\propto e^{\lambda t_b}$.  There are two
phase-space directions and two Lyapunov exponents for each mechanical degree of freedom.
{\it Sets}, indicated by braces $\{ \ \dots \ \}$, of both ``local'' [ time-dependent, indicated by $(t)$ ]
 and ``global''  [ time-averaged, indicated by $\langle \ \dots
\ \rangle $ ] Lyapunov exponents can be used to describe this chaos, with
$$
\{ \ \lambda_{\rm global} \equiv \lambda \equiv \langle \ \lambda(t) \ \rangle \equiv
\langle \ \lambda_{\rm local} \ \rangle \ \} \ .
$$
Details of this exponentially-diverging chaos became available with the advent of fast computers enabling
low-cost numerical solutions of the atomistic motion equations.  The usual procedure was, and is, to
generate a ``reference trajectory'' and one or more ``satellite trajectories'', keeping track of the
tendency of the satellite trajectories to diverge away from or approach closer to the reference\cite{b3,b4,b5}.
To distinguish this reference trajectory, $( \ q_0 \ , \ q_{dt} \ , \ q_{2dt} \ , \ \dots \ )$ from its
reverse, $( \ \dots \ , \ q_{2dt} \ , \ q_{dt} \ , \ q_{0} \ )$ we will sometimes term these the 
``primary'' and ``reversed'' coordinate sets.

The separations of the satellite trajectories from the reference define an orthogonal set of ``offset
vectors'' in the phase space, $\{ \ \delta(t) \equiv (q,p)_{\rm sat} - (q,p)_{\rm ref} \ \}$ .  The
underlying ``molecular dynamics'' simulations require five ingredients: forces, initial conditions,
boundary conditions, integrators, and diagnostics.  Good choices of these five ingredients can give
insight into the symmetries and the broken symmetries of Hamiltonian chaos.  In what follows we will
emphasize ``important'' particles, those particles making above-average contributions, $( \ \delta q^2 + 
\delta p^2 \ )$ to the offset vector which measures the most rapid divergence of the satellite trajectory
from the reference. 

Here we select two special Hamiltonian problem types: the dynamics of a single soft disk\cite{b6}
and the inelastic collision of two many-particle solid bodies\cite{b7,b8}. Our interest in the
single-particle problem is primarily pedagogical, especially for its apparent ergodicity and for the
simplicity of its offset-vector structure.  The single-disk ``cell-model'' problem has only one pair of
chaotic offset vectors, a system particularly easy to analyze.  Both problem types reveal two
interesting aspects of Hamiltonian chaos.  First the local Lyapunov exponents have a tendency to
{\it pair, corresponding to the forward-backward time reversibility} of Hamiltonian motion.
The single-disk cell model dynamics  apparently illustrates pairing {\it all} of the time, once the
transient behavior from the initial conditions has decayed.  The inelastic collision problems illustrate
pairing only {\it most} of the time.  During the collision process pairing is destroyed.

There is a second consequence of chaos present in both problem types.
These Lyapunov exponent pairs illustrate {\it symmetry breaking} --- for both types, the one-body cell-model
problem and the collisional many-body problems.  This is because the forward and backward sets of
exponent pairs ,
$$
\{ \ \pm \lambda^{\rm backward}(t) \ \} \ \leftarrow \ \{ \ q(t) \ \} \
\rightarrow \ \{ \ \pm \lambda^{\rm forward}(t) \ \} \ ,
$$
can be quite different {\it along exactly the same trajectory} (both the primary and the reversed
orderings) {\it and at exactly the same configuration}.  This difference reflects the difference between
the ``past'' and the ``future''.  From the qualitative standpoint past and future are about the same for
the one-body cell model.  Past and Future can and do differ {\it substantially} (as described by the
Second Law) for the colliding many-body systems treated here.

Demonstrating instantaneous pairing is a numerical challenge. Pairing appears to be present {\it all the
time} in the simple cell-model problem, with
$$
\lambda^{\rm forward}_i(t) = -\lambda^{\rm forward}_{5-i}(t) {\rm \ \ and \ \ } 
\lambda^{\rm backward}_{i}(t) = -\lambda^{\rm backward}_{5-i}(t) \ .
$$
On the other hand our numerical work on many-body problems shows that the tendency toward pairing can be
defeated by strong localized events.  We find that pre-collision pairing is destroyed by energetic
collisions of small crystallites, but can apparently recur as the coalesced body equilibrates.  We will
see clearly that Lyapunov-exponent pairing can be destroyed {\it during} the collision process.  We also
find that a single trajectory's stability can be quite different, forward and backward in time.  Forward
and backward stabilities, for the same configuration but reversed momenta can and do differ qualitatively.
This is a bit surprising.  If similar trajectories separate, when propagated forward in time, they correspond
to approaching trajectories in the reversed motion.  In an idealized  perfectly
time-reversible situation the first most-positive time-averaged Lyapunov exponent would correspond to the
last most-negative exponent if all the geometric data were processed ``backward'', in the opposite order.

In fact, things are not so simple. Typically $\lambda_1^{\rm forward}(t)$ doesn't correspond to {\it any} of the
backward exponents.  The exponents from a forward processing of coordinate data are not simply related to
those from a backward processing.  The many-body inelastic-collision problem clearly illustrates this
symmetry-breaking exponent pairing.  The forward and backward exponent pairs are quite different for
{\it exactly the same} configuration.  In addition there is a qualitative distinction to be seen in the
phase-space separation vectors associated with the largest (and smallest) Lyapunov exponents.  And the
offset-vector differences {\it forward} in time don't resemble those with time reversed.  These seemingly
odd differences invariably emerge when time-reversible
Hamiltonian mechanics is applied to highly nonequilibrium situations.  We will see that the ``important
particles'' going forward in time can be quite different to those in the reversed motion {\it at the same
configuration and with reversed momenta}.  This symmetry-breaking, with $\{ \ \lambda^{\rm forward} \ \}$
very different to $\{ \ \lambda^{\rm backward} \ \}$ as well as the transient nature of the pairing,
$\{ \ +\lambda \ \} = \{ \ -\lambda \ \}$ , both forward and backward, surprised us and prompted us to
write this paper.  A second motivation was the delay in publication of our manuscript [ar$\chi$iv:1112.5491]
``Time's Arrow for Shockwaves; Bit-Reversible Lyapunov and Covariant Vectors ; Symmetry Breaking'', 
submitted to the Journal of Physics A in December of 2011 and finally withdrawn and published in
Computational Methods in Science and Technology in early 2013!\cite{b9}

This paper is organized as follows.  We fix ideas by beginning with the simplest possible one-body Hamiltonian
problem.  We describe this chaotic problem in Section II, and use it to illustrate Lyapunov instability, the
forward-backward pairing of the local exponents, and symmetry breaking.  We follow a simplification suggested
by Romero-Bastida {\it et alii}\cite{b10}, using Levesque and Verlet's bit-reversible leapfrog
algorithm\cite{b11} to generate arbitrarily-long perfectly-time-reversible trajectories, both forward and
backward in time.

In Section III we consider two larger but still quite manageable problems.  In both of them we analyze the
inelastic collision of two similar cold crystals.  The minimal $N=14$-body simulation of two colliding seven-body
hexagons characterizes the stability of the motion in a 56-dimensional $\{ \ x,y,p_x,p_y \ \}$ phase space.
Describing any 14-body trajectory in that space involves solving 56 ordinary differential equations.  Evaluating the
{\it stability} of that motion (the 56-dimensional response to perturbations in 56 directions) requires
solving $56^2$ more differential equations, giving 3192 in all.  A more detailed study, following the collisions
of two $(4+5+6+7+6+5+4=37)$-particle hexagonal crystallites in their $4\times74$-dimensional phase space involves
solving $297\times 296 = 87,912$ ordinary differential equations.  These describe the motion of 296 orthogonal
296-dimensional ``offset vectors''.  The vectors are made orthonormal at the conclusion of every timestep, with a
typical collision analysis requiring a few million timesteps.

The interesting topological features connecting an inelastic collision's local Lyapunov spectrum to the
phase-space offset vectors can be illustrated by generating a {\it reference} trajectory in either of two
different ways, [1] bit-reversibly\cite{b9,b10,b11} or [2] (slightly) irreversibly, with a highly-accurate fourth-order
Runge-Kutta integration.  The excellent agreement furnished by these two quite different approaches supports the
use of both algorithms.  In either case the $4N$ orthonormalized {\it satellite} trajectories are generated with the
classic fourth-order Runge-Kutta integrator along with the Gram-Schmidt orthonormalization algorithm.

In Section IV we consider an 800-body problem, where the evolution of the inelastic-collision dynamics takes too
long (a few sound-traversal times) for accurate time-reversal using double-precision Runge-Kutta integration.
The alternative bit-reversible technique allows us to identify the ``important particles'' [above-average
contributors to $\lambda_1(t)$] for this highly irreversible process, and provides a clear distinction between
the stabilities of the forward and backward (primary and reversed) dynamics.  Section V is our Conclusion and
Summary, relating all these time-reversible model results to the {\it irreversibility} inherent in the
Second Law of Thermodynamics and to microscopic Lyapunov instability.

%Figure 1 goes here                                                                                             \
                                                                                                                 
\begin{figure}[h]
\vspace{1 cm}
\includegraphics[width=7.5cm,angle=-90]{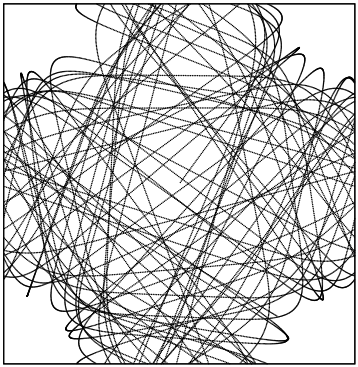}

\caption{
Sample cell-model trajectory segment for $0 \ < \ t \ <  \ 200$ and $dt = 0.001$.
}
\end{figure}

\section{One-Body Cell Model Dynamics}
This simplest chaotic problem is the dynamics of a soft Hamiltonian disk with two degrees of freedom, confined
within a periodic square lattice of similar soft-disk scatterers.  ``Cell models'' of this type were studied early
in the last century.  The corresponding one-particle partition-function models provided semiquantitative
``free-volume'' estimates for the many-body partition functions characterizing the then-somewhat-mysterious
liquid state\cite{b12}.  The dynamics for this cell-model system occupies a  three-dimensional constant-energy
volume in the four-dimensional $\{ \ x,y,p_x,p_y \ \}$ phase space.  See Figure 1 for a configuration-space
view of the dynamics.  For this problem, with its periodic boundaries, no attractive forces are necessary.
Accordingly, we use a purely-repulsive potential energy (with numerical integration errors
minimized by choosing a pair potential with three continuous derivatives at the cutoff distance of unity) :
$$
\phi(r<1) = (1 - r^2)^4 \ \longrightarrow \ F(r<1) = 8r(1 - r^2)^3 \ .
$$
Punctuation of the free-flight regions, by very smooth collisions, enhances the accuracy of the numerical work.
For definiteness (so that a diligent reader can reproduce our results in detail) the initial velocity is
(0.6,0.8) with the initial coordinates (0,0) in the center of a square-lattice periodic cell. The periodic
boundary conditions, $( \ -1 < x,y < +1 \ )$, are imposed by adding or subtracting, if necessary, the cell
width 2 at the end of every timestep.  Because the spacing between the centers of the fixed nearest-neighbor
scatterers is 2, the moving particle interacts with at most one of the fixed particles.  For definiteness
we choose the initial four offset vectors parallel to the four Cartesian phase-space directions:
$(x,y,p_x,p_y) \ $.  With fourth-order Runge-Kutta integration, the calculation is insensitive to changes of the
timestep, $dt = 0.001$, and the length of the offset vectors, $|\delta | = 0.00001 \ $.  The results
described below are obtained by following the dynamics of {\it five} separate trajectories, the
``reference'' trajectory along with four nearby ``satellite'' trajectories,
with the differences defining the four offset vectors $\{ \ \delta_1 \dots \delta_4 \ \}$ .

To avoid the divergence of the offset vectors that would accompany exponential growth it is usual either
[1] to rescale them\cite{b3,b4} or [2] to measure their {\it virtual} rates of increase\cite{b5}, which
can be expressed in terms of Lagrange multipliers constraining satellite trajectories to remain at a
fixed separation from a reference trajectory.  Additional multipliers constrain the directions of the
satellite trajectories to remain orthogonal.  Numerical work indicates that the positive Lyapunov exponent
$\lambda_1^{\rm forward}(+t)$ is accurately paired to its mostly-negative twin $\lambda_4^{\rm forward}(+t)$ .
Typically this pair of instability exponents, forward in time, is not at all similar to the corresponding
pair of ``reversed'' or ``backward'' exponents $\{ \ \lambda_{1 \ {\rm and} \ 4}^{\rm backward}(-t)$ \ \},
 if the same coordinate trajectory is followed ``backward'' in time.

%Figure 2 goes here                                                                                             \

\begin{figure}[h]
\vspace{1 cm}
\includegraphics[width=7.5cm,angle=-90]{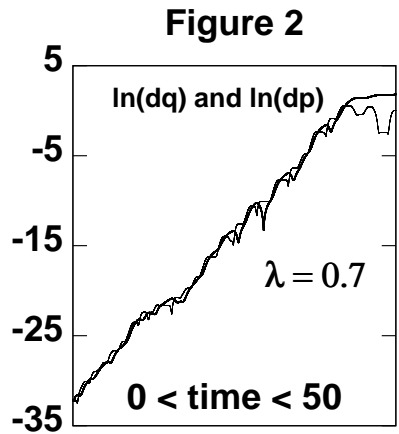}

\caption{
Growth of small perturbations in the coordinate $q$ and momentum $p$ with time.
}
\end{figure}

Despite the pairings, $\lambda_1 + \lambda_4 \simeq \lambda_2 + \lambda_3 \simeq 0$ , the primary exponents,
 measured in the forward time direction, reflect the past rather than the future.  The reversed exponents,
measured for the other ``backward'' time direction, are different.  The backward exponents anticipate the
``future'' rather than reflecting the past.  For a typical numerical trajectory segment, which can be
followed either forward or backward, see Figure 1.  In  Figure 2 we see the near-perfect
exponential divergence of a small perturbation ,
$$
(x,y)_{t=0} = (10^{-16},0) \ .
$$
The slope gives an estimate of the largest Lyapunov exponent, $\lambda_1 \simeq 0.7$.  By symmetry
the {\it smallest} exponent is $-0.7$ so that the time-averaged Lyapunov spectrum is
$$
\{ \ \langle \ \lambda \ \rangle \ \} = \{ \ +0.7, \ 0.0, \ 0.0, \ -0.7 \ \} \ ,
$$
in the four-dimensional phase space of the Hamiltonian motion.

The simplest algorithm characterizing the disk's Lyapunov instability in this
space follows the dynamics of a single time-reversible reference trajectory along with
 four nearby satellite trajectories.  The reference-to-satellite vectors $\{ \ \delta_i \ \}$
are constrained to remain orthogonal at the end of each timestep, maintaining the constant length
$\delta \equiv 0.00001$.  The Gram-Schmidt orthonormalization algorithm first rescales $\delta_1$
and then removes the projection of $\delta_2$ in the direction of $\delta_1$ :
$$                                                                                                                             
\delta_2 \longrightarrow \delta_2 - \delta_1 [ \ \delta_1 \cdot \delta_ 2 \ ]/\delta^2 \ .                                     
$$
The rescaling operation gives the local value of the Lyapunov exponent $\lambda_1$ :
$$                                                                                                                             
\lambda_1(t) = (1/dt)\ln ( \ \delta_1/\delta \ ) \ .                                                                           
$$
Then $\delta_2$ is rescaled [giving the second local Lyapunov exponent $\lambda _2(t)$]
and the projections of $\delta_3$ in the directions of $\delta_1$
and $\delta_2$ are removed :
$$                                                                                                                             
\delta_3 \longrightarrow \delta_3 - \delta_1 [ \ \delta_1 \cdot \delta_3 \ ]/\delta^2                                          
- \delta_2 [ \ \delta_2 \cdot \delta_ 3 \ ]/\delta^2 \ .                                                                       
$$
Finally $\delta_3$ is rescaled, giving $\lambda_3(t)$ and $\delta_4$ is similarly made
orthogonal to $\{ \ \delta_1,\delta_2,\delta_3 \ \}$ and rescaled to give $\lambda_4(t)$.
In the end four orthogonal vectors $\{ \ \delta_i(t) \ \}$ and four local Lyapunov exponents
$\{ \ \lambda_i(t) \ \}$ result.  

%Figure 3 goes here                                                                                             \

\begin{figure}[h]
\vspace{1 cm}
\includegraphics[width=7.5cm,angle=-90]{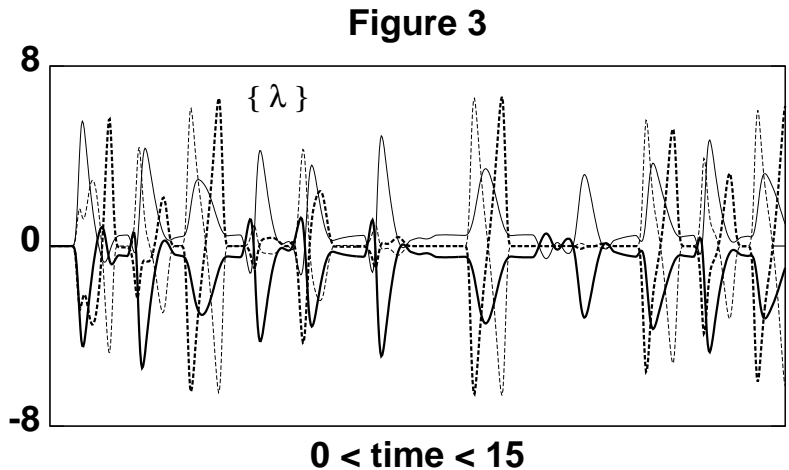}

\caption{
The four local Lyapunov Exponents for the cell model illustrating exponent ``pairing''.
}
\end{figure}

After a transient time of order $( \ 1/\lambda_1 \ )$ both the sets of Lyapunov vectors
(forward and backward) for the cell model are observed visually to ``pair'', with about
six-figure numerical accuracy :
$$                                                                                                                             
\lambda_1(t) + \lambda_4(t) \simeq \lambda_2(t) + \lambda_3(t) \simeq  0 \ .                                                             
$$
See  Figure 3 for a sample sequence obtained with Runge-Kutta timestep $dt=0.002$ and with
the orthogonal vector length $|\delta| = 0.000001$.  Again, the results are not at all sensitive
to either of these choices.
Because Hamiltonian mechanics is strictly time-reversible, with all the rates changing sign in a
time-reversed simulation, {\it exact}  pairing, as suggested by our numerical cell-model results,
is certainly a {\it plausible} property of cell-model trajectories.  Nevertheless, we will be
considering three other chaotic Hamiltonian systems which clearly violate this pairing property,
at least some of the time, in the next two Sections.

There is a set of first-order ordinary differential equations equivalent to the Gram-Schmidt
procedure just described in the small timestep limit\cite{b5}, $dt \longrightarrow 0$ :
$$                                                                                                                             
\dot \delta _1 = D\cdot \delta_1 - \lambda_{11}\delta_1 \ ;                                                                    
$$
$$                                                                                                                             
\dot \delta _2 = D\cdot \delta_2 - \lambda_{21}\delta_1 - \lambda_{22}\delta_2   \ ;                                           
$$
$$                                                                                                                             
\dot \delta _3 = D\cdot \delta_3 - \lambda_{31}\delta_1 - \lambda_{32}\delta_2  - \lambda_{33}\delta_3  \ ;                    
$$
$$                                                                                                                             
\dot \delta _4 = D\cdot \delta_4 - \lambda_{41}\delta_1 - \lambda_{42}\delta_2  - \lambda_{43}\delta_3  - \lambda_{44}\delta_4\
  \ .                                                                                                                          
$$
Here the matrix $D$ describes the effect of the perturbations $\{ \ \delta \ \}$ on the {\it unconstrained}
motion of the vectors.  The ten Lagrange multipliers $\{ \ \lambda_{i\ge j} \ \}$ vary with time so as to
maintain the ten orthonormality constraints,
$\{ \ \delta_i\cdot \delta_j \equiv \delta^2\delta_{ij} \ \}$ .
The diagonal Lagrange multipliers in these differential equations are identical to the local Lyapunov
exponents, $\lambda_{ii} \equiv \lambda_i(t)$.  It is easy to show that the differential
equations are perfectly time-reversible (in the sense that the coordinates are unchanged while the
momenta and Lagrange multipliers change sign).  This apparent but illusory time symmetry is {\it broken},
even for simple systems such as our one-particle cell model.  It is also easy to show that {\it exactly
the same} ten Lagrange multipliers result if the basis vectors are used to describe the virtual growth
rates of a two-trajectory {\it length}, a three-trajectory {\it equilateral triangle}, and a
four-trajectory {\it regular tetrahedon}.

For relatively short times solutions of this simple dynamical system can be generated with Runge-Kutta
integration.  The longtime irreversibility of such Runge-Kutta integrations is due to the cumulative growth
of single-timestep errors.  These local errors are proportional to $dt^5$ times the fifth time derivative
of the phase-space variables.  To avoid the resulting longtime irreversibility the dynamics can instead be
generated as an ordered series of coordinate values $\{ \ ( \ x_t,y_t \ ) \ \}$ using a somewhat less
accurate but completely ``bit-reversible'' integer algorithm for the reference trajectory.  Among them,
Levesque and Verlet's third-order algorithm\cite{b10,b11} is certainly the simplest :
$$
\{ \ q_{t+dt} - 2q_t + q_{t-dt} \equiv [ \ F_tdt^2/m \ ]_{\rm Integer} \ \} \ .
$$
Rather than the phase variables $\{ \ q_t,p_t \ \}$ two sets of adjacent coordinate values
$\{ \ q_t,q_{t\pm dt} \ \}$ are required to start the Levesque-Verlet algorithm.
Here the coordinates and their second differences are all evaluated as (large) integers.  The resulting
bit-reversible reference trajectory can be extended infinitely far into the future or the past
without any need to store the trajectory.
A set of momenta corresponding to the coordinates ,
$$
\{  \ \dots \ , \ p_{t-dt} \ , \ p_t \ , \ p_{t+dt}  \ , \ \dots \ , \ \} \ ,
$$
and, like the coordinates, with third-order accuracy in $dt$ , can be defined as follows\cite{b9} :
$$
p_t \equiv ( \ 4/3 \ )[ \ q_{t + dt} - q_{t - dt}  \ ]/( \ 2dt \ ) -
                  ( \ 1/3 \ )[ \ q_{t +2dt} - q_{t - 2dt} \ ]/( \ 4dt \ ) \ .
$$

The nearby satellite trajectories are generated with the usual Runge-Kutta integration.  By using 16-byte
integers the accuracy of the integer-algorithm's reference trajectory can be made to match that of a
double-precision floating-point simulation.

A practical approach uses bit-reversible integration for the reference trajectory and fourth-order
Runge-Kutta integration for the four nearby satellite trajectories.  At the end of each timestep
we use Gram-Schmidt orthonormalization, keeping the lengths of the four ``offset vectors'' fixed
$\{ \ |r_s - r_r| = \delta \ \} \ $ and their directions orthogonal.  The accuracy
of the Lyapunov spectrum depends (relatively weakly) upon the timestep $dt$ and the vector length
$\delta $ .  A convenient initial condition ,
$$
\{ \ x, \ y, \ p_x, \ p_y \ \} = \{ \ 0.0, \ 0.0, \ 0.6, \ 0.8 \ \} \ ,
$$
with total energy $E = K + \Phi = ( \ 1/2 \ ) \ge \sum \phi$ , guarantees that the moving particle can
get no closer to any of its four fixed neighbors than a distance
$r_{\rm min} = \sqrt{(1 - (1/2)^{1/4})} = 0.3988779$.
At the end of each timestep the periodic boundary conditions are applied to ensure that the moving disk
stays within its periodic cell.  A million timestep simulation using the classic fourth-order Runge-Kutta
integrator for the reference trajectory  with $dt = 0.0002$ exhibits an energy loss less than one part
in $10^{13}$ .

Long time energy loss can be avoided entirely, and the numerical trajectory can be made precisely
time-reversible, by using Levesque and Verlet's bit-reversible integrator.  That algorithm requires a
pair of subroutines mapping the floating-point interval $\{ \ -2 < {\tt float} < +2 \ \}$ onto the
integer interval $\{ \ -M < {\tt integer} < +M \ \}$ :
$$
{\tt integer  = float*M/2.0d00} \longleftrightarrow   {\tt float = 2.0d00*integer/M} \ .
$$
We choose $M=10^{16}$ so that the precision of the bit-reversible simulation is comparable
to that of a typical double-precision fourth-order Runge-Kutta simulation.

%Figure 4 goes here                                                                                             \

\begin{figure}[h]
\vspace{1 cm}
\includegraphics[width=7.5cm,angle=-90]{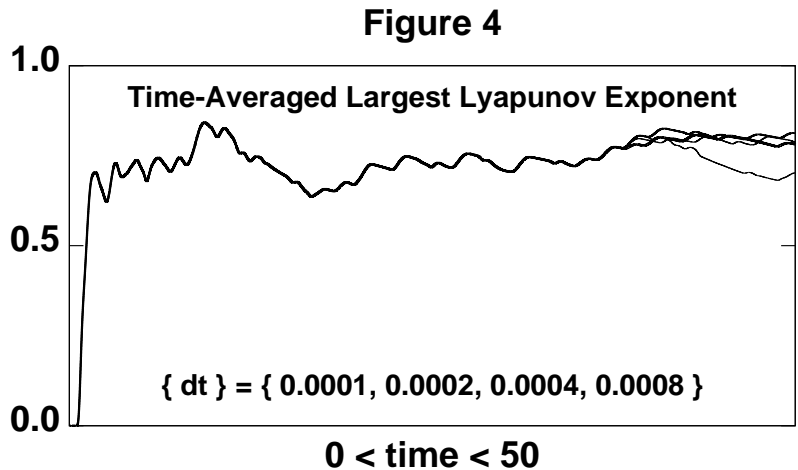}

\caption{
Dependence of the local Lyapunov exponent on the timestep $dt$.
}
\end{figure}

Two initially different offset vectors come to agree --- to six-figure accuracy --- after propagating for
a time of order
20.  Figure 4 shows the early stages of convergence of the nonzero Lyapunov exponent for four different
timesteps, 0.0001, 0.0002, 0.0004, and 0.0008.  From the visual standpoint the results are identical until a
time of order 40, where Lyapunov instability causes the four trajectories to separate.  This long-time-averaged
Lyapunov spectrum, $\{ \ +0.7, \ 0, \ 0, \ -0.7 \ \}$ is, as we would expect,  perfectly consistent  with the
two-system offset vector calculation documented in  Figure 2.

The more complicated simulation of Figure 3, giving the whole spectrum, involves solving 20 ordinary
differential equations -- four of them describing the reference trajectory and 16 more describing its four
satellite trajectories.  This one-body cell-model problem is an excellent warmup exercise for the many-body
problems described in what follows in the next two Sections.  These upcoming many-body examples are more
complex, in that exponent pairing is a {\it transient} (and therefore only approximate) phenomenon.  The
loss of pairing is evidently associated with dynamical events that appear irreversible, brought about by
the choice of inhomogenous out-of-the-ordinary initial conditions.

\section{Inelastic Collisions of Two Cold Hexagonal Crystallites}
{\it Thermodynamic} irreversibility occurs whenever mechanical energy is dissipated into heat. We wish
to see how such thermodynamic irreversibility is reflected in the Lyapunov instability of atomistic
simulations of conservative Hamiltonian mechanics.   To begin we will consider a simple demonstration
of irreversible behavior, the inelastic collision of two cold seven-atom crystallites to form a single
hotter 14-body drop.  Our first experience with this general problem type, in 1990, was intended to measure
the ``coefficient of restitution'' for two bouncing balls.  But the balls refused to bounce, instead
fusing, so as to form a single ball, just as in the present work.  The earlier two-ball work is
mentioned, and illustrated, in Reference 7.  A recent four-ball analog appears on page 96 of
(the second [2012] edition) of Reference 1.
This same combination of the many-body embedded-atom potential with the repulsive core potential is
useful for modeling surfaces and other lattice defects, as well as the dynamics of plastic flows\cite{b13}.
For the problems considered here the vapor pressure of the coalesced balls is so low that no special spatial
boundary conditions are required to contain all the particles.

Each particle has unit mass.  In addition to the repulsive pair forces derived from the $(1-r^2)^4$ pair
potential, we add on a longer-range attractive smooth-particle potential based on the deviations of the
individual particle densities from unity, as calculated from Lucy's smooth-particle weight function\cite{b1},
with a range $h=3.5$:
$$
\Phi (\{ \ \rho \ \}) \equiv \sum_{i=1}^{14} (1/2)(\rho_i - 1)^2 \ ; \
\rho_i = \sum_{j=1}^{14} w( \ | \ r_i -r_j \ | \ ) \ ;
$$
$$
w_{\rm Lucy}( \ r<h=3.5 \ ) = ( \ 5/\pi h^2 \ )[ \ 1 + 3z \ ][ \ 1 - z \ ]^3 \ ; \ z \equiv ( \ r/h \ ) \ .
$$
Lucy's weight function is normalized to reflect the local density, with
$$
\int_0^h2\pi rw(r) \equiv 1 \ .
$$
The contribution of the smooth-particle potential to the equations of motion is
$$
\ddot r_i = \sum_j[ \ ( \ 1 - \rho_i \ )\nabla_iw_{ij} + ( \ 1 - \rho_j \ )\nabla_iw_{ij} \ ]
= \sum_j( \ 2 - \rho_i - \rho_j \ )\nabla_iw_{ij} \ .
$$
 Figure 5 shows a series of snapshots of two colliding 7-particle hexagons with the time reversed
at $t = 100$.

%Figure 5 goes here                                                                                             \

\begin{figure}[h]
\includegraphics[width=9cm,angle=-90]{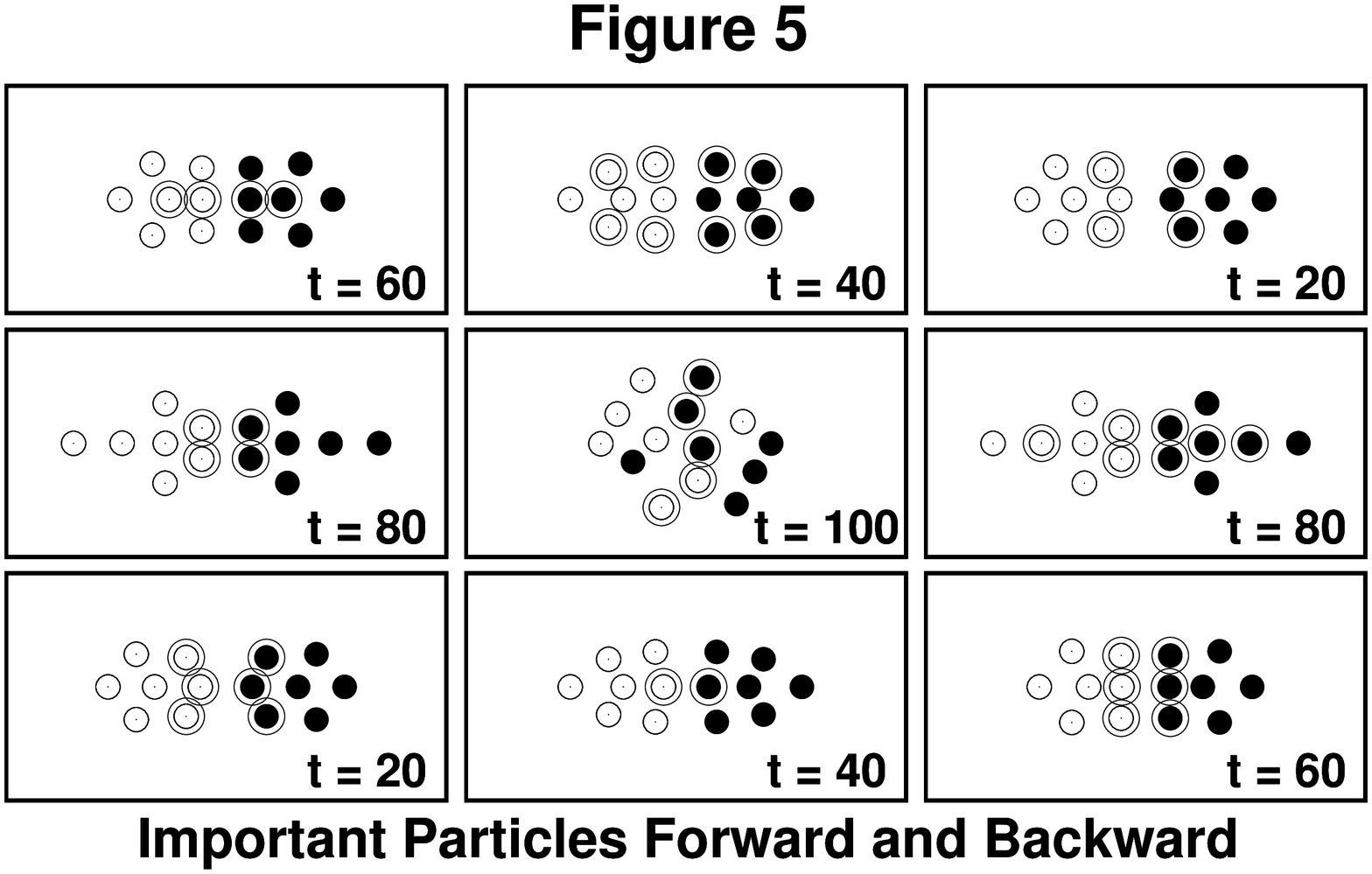}

\caption{
Important particles in a 14-particle inelastic collision time-reversed at $t=100$.
}
\end{figure}

%Figure 6 goes here                                                                                             \

\begin{figure}[h]
\includegraphics[width=9cm,angle=-90]{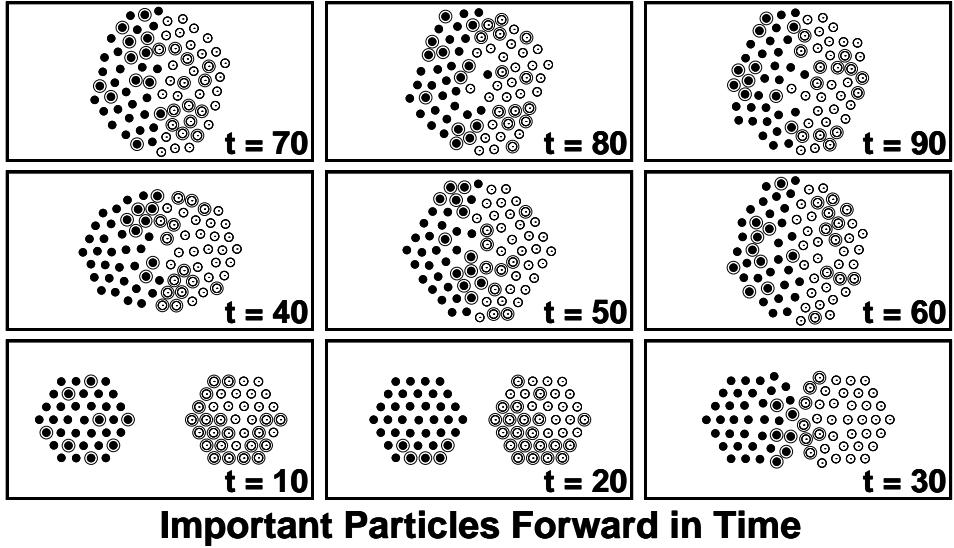}

\caption{
Important particles are emphasized in the collision of two 37-particle crystallites.
}
\end{figure}

%Figure 7 goes here                                                                                             \

\begin{figure}[h]
\includegraphics[width=9cm,angle=-90]{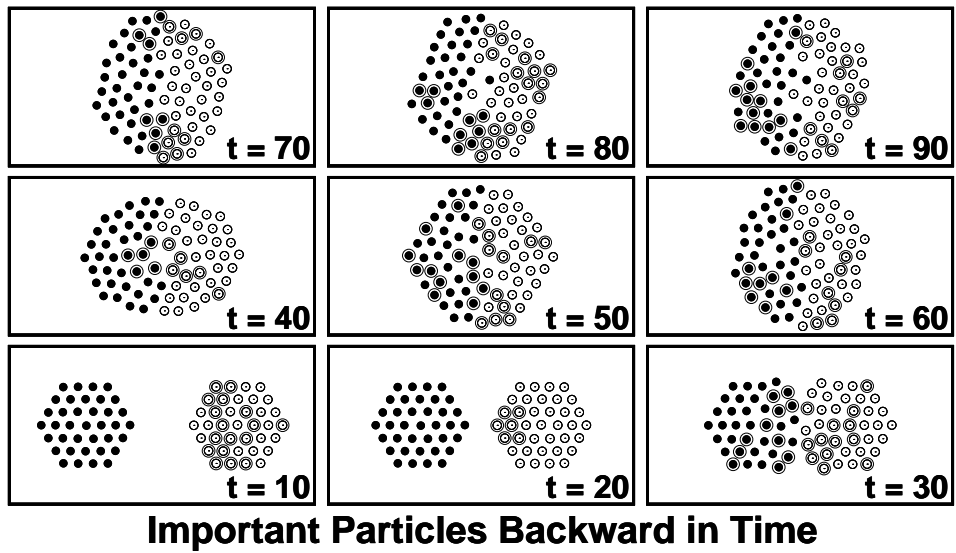}

\caption{
Same as Figure 6 but with the bit-reversible trajectory processed backward in time.
}
\end{figure}

 Figures 6 and 7 show similar series of snapshots for two 37-particle hexagons.  Just as before, the
initial velocities are $p_x = +0.1$ for those particles in the left hexagon
and $p_x = -0.1$ for those at the right.  In these figures particles making an
above-average contribution to the local Lyapunov exponent $\lambda_1(t)$ are distinguished by an extra
circular ring for emphasis. Note particularly that in the forward-in-time motions the leading-edge
particles contribute most to instability.  In the reversed collision, with the drop separating into two
hexagons, the cooperative motion of the interior particles is more important to the stability.
In the initial least-energy cold configuration for Figure 5 the nearest-neighbor spacing is
0.8611 2127 0463 and the seven-body crystal's comoving energy is 0.6390 2960 9388.  The energy is
positive due to the contribution of the attractive potential, which
vanishes at a density of unity, not zero.  In Figures 6 and 7 we have chosen a stronger repulsive pair
potential, $10(1-r^2)^4$ rather than $(1-r^2)^4$, in order to compensate somewhat for the layering tendency of
the embedded-atom interaction.

We began by investigating such two-hexagon collisions with the classic fourth-order Runge-Kutta integrator.
Although the {\it energy} changes can be made negligible for elapsed times of several hundred, Lyapunov
instability eventually spoils the details of a ``reversed'' Runge-Kutta trajectory, and in a much shorter
time, of order 25.  See again Figure 4. Quadruple precision would simply double this time, to 50. 
Energy conservation provides no hint of this trajectory irreversibility.  Choosing a timestep of $dt = 0.001$
conserves the energy to an accuracy of twelve digits over the course of a 600,000 timestep run.  But the
time reversibility is effectively destroyed much sooner, at about 25,000 timesteps.

To maintain precise time reversibility in our Lyapunov computations, we used the Levesque-Verlet
bit-reversible integrator.   Figures 5-7 are based on bit-reversible reference trajectories with
Runge-Kutta satellite trajectories orthonormalized at each timestep.
For the same number of force evaluations per unit time the bit-reversible timestep could be made four times
smaller:
$$
dt_{\rm bitrev} = (1/4)dt_{\rm RK4} = 0.00025 \ .
$$
But for simplicity we have used $dt = 0.001$ for both integrators.

%Figure 8 goes here                                                                                             \

\begin{figure}[h]
\vspace{1 cm}
\includegraphics[width=8cm,angle=-90]{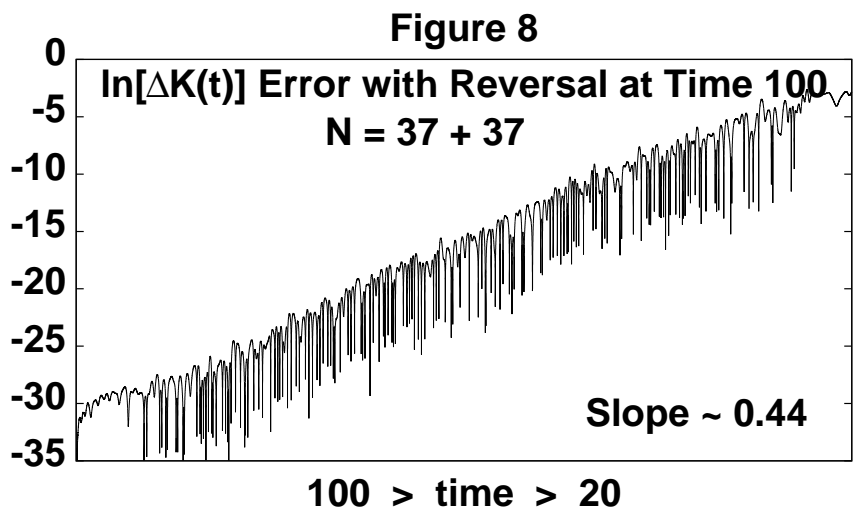}

\caption{
Exponential growth of kinetic energy error after time-reversal at time 100.
}
\end{figure}

%Figure 9 goes here                                                                                             \

\begin{figure}[h]
\vspace{1 cm}
\includegraphics[width=8cm,angle=-90]{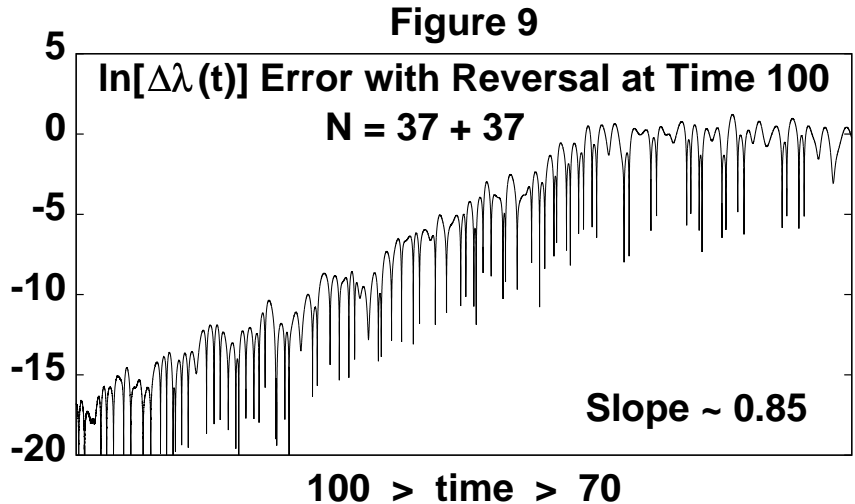}

\caption{
Exponential growth of Lyapunov exponent error after time-reversal at time 100.
}
\end{figure}

%Figure 10 goes here                                                                                             \

\begin{figure}[h]
\includegraphics[width=8.5cm,angle=-90]{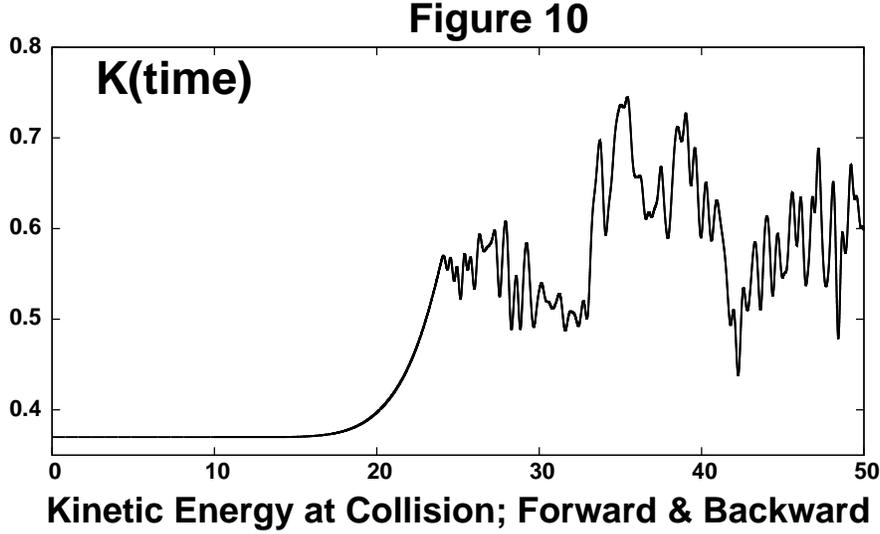}

\caption{
Variation of Kinetic energy (initially $37\times2\times0.005$) during the bit-reversible
inelastic collision of two 37-particle crystallites.
}
\end{figure}

%Figure 11 goes here                                                                                             \

\begin{figure}[h]
\vspace{1 cm}
\includegraphics[width=10.0cm,angle=-90]{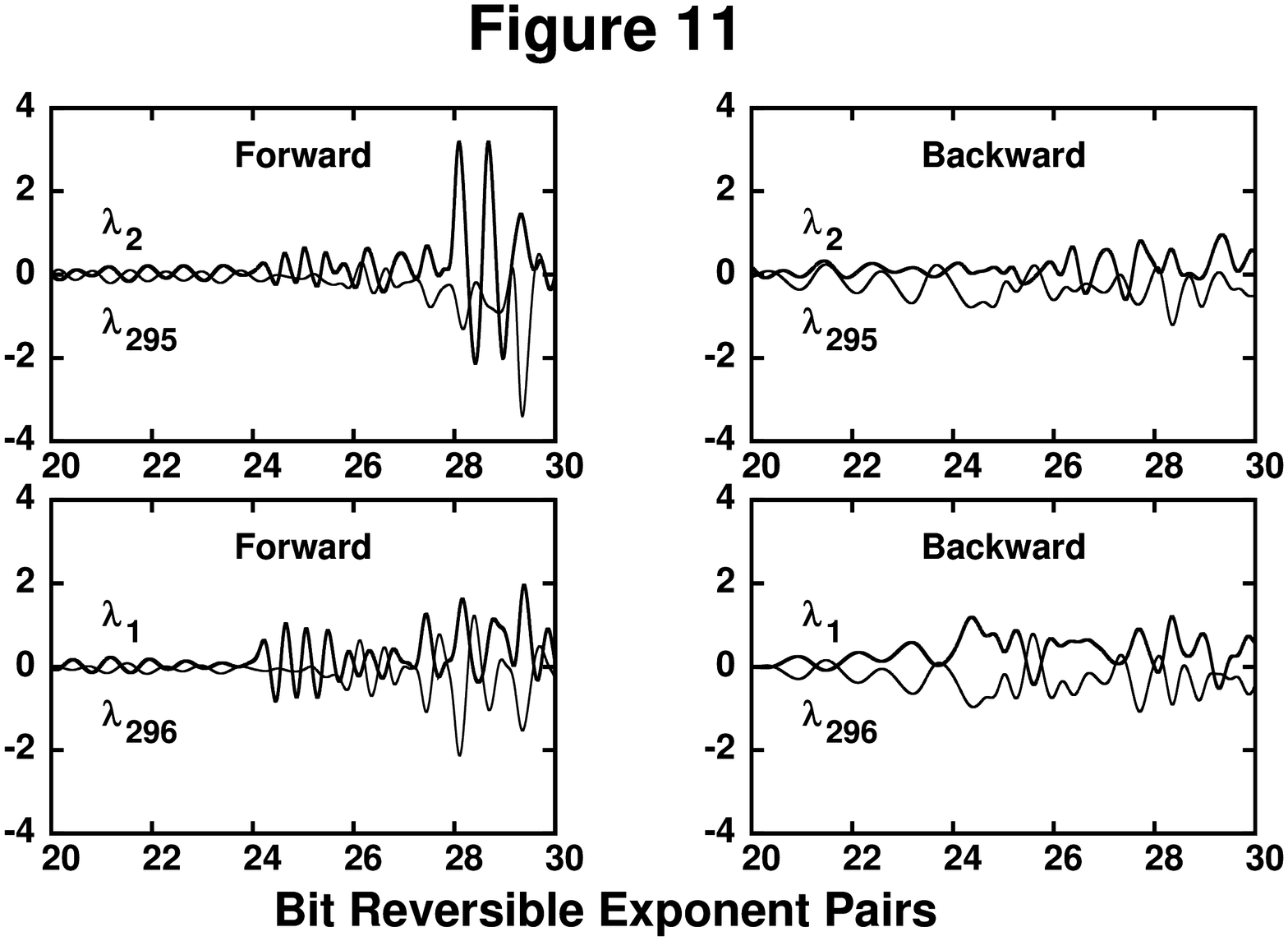}

\caption{
Lack of pairing relationships between the first and last Lyapunov exponents during
the bit-reversible simulation of a 74-body inelastic collision.
}
\end{figure}

Figures 8 and 9 compare the Runge-Kutta and bit-reversible calculations of energy and the largest Lyapunov
exponent for the $37+37$ particle problem.  Note again that the local Lyapunov exponent is a much more sensitive test
of trajectory accuracy than is the energy.  The comparison also shows that either algorithm, Runge-Kutta or
bit-reversible, can be used for simulations that are not too long. Figure 10 shows the thermalization of the kinetic
energy as the two hexagons merge to form a warm ball.  At about time 30 the coalescence is complete.  The remaining
dynamics consists of relatively featureless thermal motion.  In Figure 11 we show a portion of the
time-dependence of the 1-296, and 2-295  pairs of local Lyapunov exponents, both forward and backward in time.
From the visual standpoint simulations using a bit-reversible reference trajectory are indistinguishable from those
using Runge-Kutta integration, with time reversed, $+dt \rightarrow -dt$, at a time of 25.  These results show
very clearly that pairing is {\it not} a general phenomenon.  The more negative exponents react earlier, and more
strongly, to the collision process than do the more positive ones.

During the progress of the collision we can locate the ``important'' particles, those making above average
contributions to the length of the instability offset vector $\delta_1(t)$ .  As one might expect, the 
particles on the leading edges of the crystallites are the first to feel the collision.  In the time-reversed
motion {\it other} particles become important.  This is interesting!  We will detail this lack of time symmetry
in a larger and more complex coalescence problem in the next Section.

\section{Inelastic Collision of Two Larger Crystallites}

In two dimensions problems with a few hundred particles are already large enough to suggest continuum
flows.   Figures 12 and 13 show a series of forward and reversed snapshots from the collision of two
cold 400-particle
crystallites with the same repulsive pair potential and the same attractive embedded-atom potential as in the
74-particle problem of the last Section.  The initial state uses two copies of a 400-particle crystallite
generated by the relaxation of a $20\times 20$ square structure.  The relaxation providing initial conditions
for all these problems is easily carried out by including viscous forces, $\{ \ -(p/\tau) \ \}$ , in the
dynamics.  For simplicity, coordinates and velocities for a second crystallite were chosen to satisfy
inversion symmetry relative to the first :
$$
\{ \ x^{\rm left}(i) + x^{\rm right}(i) = 0 = y^{\rm left}(i) + y^{\rm right}(i) \ \} \ ; \
$$
$$
\{ \ p_x^{\rm left}(i) = +0.10 \ ; \ p_ x^{\rm right}(i) = -0.10 \ ;
   \ p_y^{\rm left}(i) =  0.00 = p_y^{\rm right}(i) \ \} \ . 
$$

Just as in the smaller cases this 800-particle problem exhibits two different local Lyapunov spectra,
one going forward in time and the other going backward.  The ``important particles'' are indicated
by central dots in the figures.  Here the reference trajectory is bit-reversible so that the
forward and backward particle coordinates agree to the very last bit.  The local exponents and vectors
at a time $t$ can be determined accurately by analyzing the trajectory segment from $t-30$ to $t+30$ .

%Figure 12 goes here                                                                                             \

\begin{figure}[h]
\includegraphics[width=8.5cm,angle=-90]{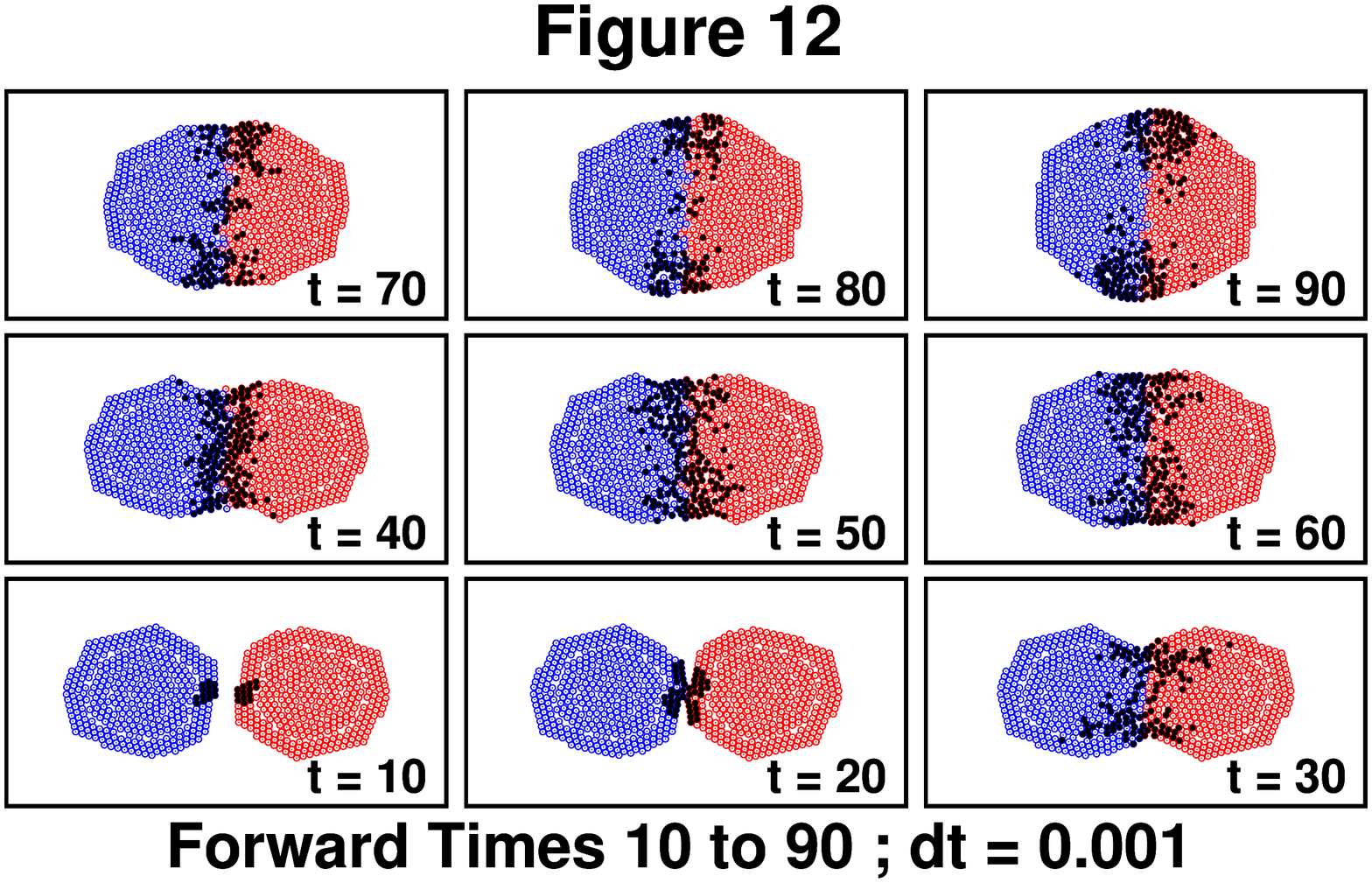}

\caption{
Important particles (black) during the collision of two 400-particle crystallites.
}
\end{figure}

%Figure 13 goes here                                                                                             \

\begin{figure}[h]
\includegraphics[width=8.5cm,angle=-90]{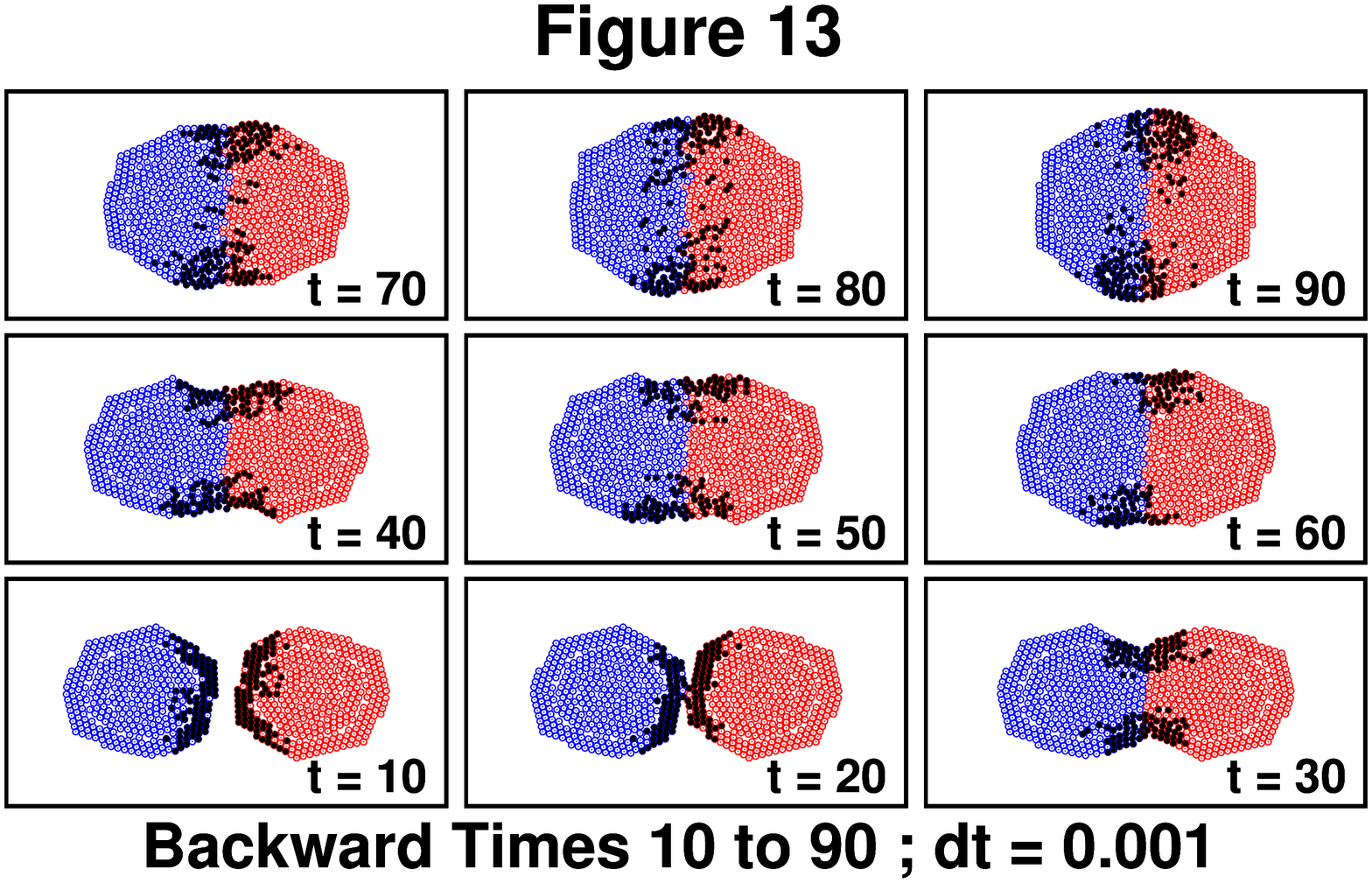}

\caption{
Important particles during the (bit-reversibly reversed) collision of Figure 12.  Note the qualitative
difference to Figure 12 with precisely identical coordinates at corresponding times.
}
\end{figure}

Figure 12 shows that forward in time the important particles are located in the collision region, where
the two crystallites first deform.  Backward in time (Figure 13) a complex collective synchronized motion of
the crystallites is required to regain the zero-temperature structures.  This ``unlikely'' motion is
localized in the necking region of the coalesced crytals.  This symmetry-breaking provides an ``Arrow
of Time'' for the coalescence problem.  The geometric features of the Lyapunov instability, given by the
offset vectors, are qualitatively different in the forward and reversed time directions.

This same symmetry-breaking is exactly the same for the recently-popularized ``covariant vectors''\cite{b10},
which are a modified approach to describing phase-space instability.  The first and last covariant
vectors correspond to the forward and reversed versions of our $\delta_1(t)$ vectors.  {\it Prediction}
of symmetry-breaking, both of the positive and negative members of each exponent pair, as well as the
symmetry-breaking distinguishing the offset vectors forward and backward in time, requires a nonlinear
analysis, as all of the equations for the reference and satellite trajectories are strictly time-reversible.

\section{Conclusion and Summary}

The observed {\it irreversibility} of simple nonequilibrium processes includes many examples from gas
theory as well as both transient and steady flows of condensed matter.  Our coalescence problems are good
examples of irreversible processes.  Deterministic time-reversible microscopic models are available to
simulate many such problems.  How does time-reversible microscopic mechanics give rise to this variety of
{\it irreversible} nonlinear macroscopic behavior?

Boltzmann's H Theorem answers this question for dilute gases\cite{b14}.  He showed that the Maxwell-Boltzmann
velocity distribution is the overwhelmingly probable result of ``uncorrelated'' collisions, collisions with
randomly-chosen impact parameters.  The thermostatted forms of reversible mechanics developed in the 1970s and
1980s provided a different explanation\cite{b8}, useful for understanding condensed matter simulations of
nonequilibrium steady states.  With the new forms of mechanics the irreversibility of nonequilibrium flows
could be traced to their extreme (fractal) rarity and to their stability, relative to their time-reversed
twins, in phase space.  Thus the {\it entropy} of {\it nonequilibrium} macroscopic states, as measured by
the (logarithm of) the number of corresponding microscopic phase-space states, is both singular and
divergent\cite{b8}  This fractal character is well-established for many simple model system\cite{b1,b2}.
In modelling a typical stationary time-reversible flow (like thermostatted plane Couette
flow or steady Fourier heat conduction) a fractal attractor forms in phase space, with a negative Lyapunov sum
giving the exponential rate of phase-volume collapse.  The time-reversed repellor, with its {\it unstable}
(positive) Lyapunov sum, provides the {\it source} for phase-space probability flow to the fractal {\it sink},
a strange attractor.  The fractal nature of such flows corresponds to the extreme rarity of nonequilibrium
steady states.  All such thermostatted simulations require a nonHamiltonian dynamics in order to generate and
account for the concentration of phase-space probability on a fractal.  

The present examples are quite different.  There are neither statistical collisions nor fractal
distributions, though there is certainly a coarse-grained macroscopic entropy increase, invisible
according to Liouville's Theorem, from minus infinity in the cold crystallites, to a positive equilibrium
value in the resulting equilibrated coalesced state.  Where does Time's Arrow come in?  The futures and
the histories of the forward (or primary) and reversed flows are (almost) exactly the {\it same} from
the standpoint of configurations $\{ \ q \ \}$.  The ``almost'' reminds us of the difficulty in constructing
a primary trajectory in the direction that violates the Second Law!  Regardless, two kinds of pairing, [1]
with any positive Lyapunov exponent paired to a corresponding negative one, and [2] with any forward
Lyapunov exponent paired to a corresponding backward one, are both consistent with the time-reversible
Hamiltonian equations of motion.  But the stabilities of the time-reversed motion equations are complicated,
in their model dependence and in their time dependence.  For flows which are relatively simple, like the
cell model, the motions in the two time directions can fail to distinguish the Lyapunov instability's
dependence on the past from its symmetry with the future.  More complex flows, like the colliding
crystallites, or shockwaves, come instead to reflect the past rather than the future.  In these cases
knowledge of $\delta_1$ automatically gives the direction in which the flow is developing.

We have seen that the Lyapunov instabilities inherent in the dynamics always reflect the past rather
than the future.  The delay between cause and effect is the same as that observed in atomistic shockwave
simulations where the stress lags the strainrate and the heat flux lags the temperature gradient\cite{b9,b15}.
The forward-backward symmetry of the microscopic motion equations does not carry through to the macroscopic
diagnostics of the motion.

Although the dynamics {\it is} symmetric in the time the {\it stability} of that dynamics is not.  The
morphology of the exponents provides a clue as to whether or not we are looking at an equilibrium system.
Whenever the past is quite different to the future this lack of symmetry can be seen in the local Lyapunov
spectrum.  The lack of pairing and the inhomogeneity of the local Lyapunov exponents needs to be related
to macroscopic entropy production.  Liouville's Theorem shows that the Lyapunov spectrum, which sums to
zero with Hamiltonian mechanics, is inconsistent with macroscopic entropy change.  On the other hand
systems like our colliding crystallites, manifesting a failure of the past and future to pair, may come to
suggest new metrics for the separation from equilibrium and its evolution. 

\section{Acknowledgments}
We are particularly grateful to Kris Wojciechowski for stimulating this work and to Franz
Waldner, Marc Mel\'endez Schofield, Vitaly Kuzkin, and Harald Posch for their patient comments and suggestions.

\end{document}